**Reporting activities for the oxygen evolution reaction: Do we compare apples to apples?**

Marcel Risch*

Nachwuchsgruppe Gestaltung des Sauerstoffentwicklungsmechanismus, Helmholtz-Zentrum Berlin, Hahn-Meitner Platz 1, 14109 Berlin, Germany

marcel.risch@helmholtz-berlin.de

**Preface**

The oxygen evolution reaction (OER) is a key enabler of sustainable chemical energy storage. Here, the author assesses the current status of protocols for benchmarking the OER in materials- and device-centered investigations and makes suggestions for more comparable data.

**Text body**

Sustainable, climate friendly, alternatives to fossil resources are needed to meet the needs of the energy and chemical sectors. Precursor feeds of non-potable water[1] and in some cases also aqueous nitrogen or carbon dioxide could be electrochemically reacted to sustainably produce many key fuels and valuable chemicals from renewable sources using in devices such as electrolyzers or photoelectrochemical cells (Figure 1). The water oxidation reaction (WOR) or oxygen evolution reaction (OER) at the anode takes a pivotal role in this approach as it provides the protonated ions for the reduction of precursors at the cathode to the desired fuel or chemical. Four electrons and ions need to be transferred to make $O_2$ from $2H_2O$ (or $4OH^-$ in alkaline media), at the cost of large overpotential at the fuel- or chemical-producing cathode. Additionally, the needed high potentials to drive the OER may degrade various electrode components.[2] Therefore, the identification of stable and active electrocatalysts for the OER has received considerable interest in the last decades. While stability and activity are both important and usually linked, this Comment focusses on reporting the *activity* of OER electrocatalysts in the context of the most mature field of water electrolysis and in particularly the issue of comparability among reported activities in current benchmarking studies.

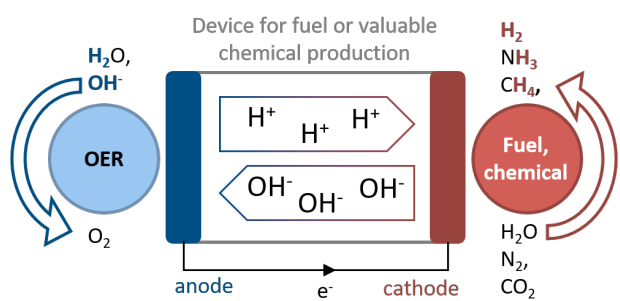

**Figure 1. Significance of the oxygen evolution reaction (OER).** The OER at the anode of devices for the production of fuels or valuable chemicals provides the ions needed for the reduction of precursors such as water, nitrogen or carbon dioxide to the desired fuel or chemical.

Bligaard et al.[3] define benchmarking in catalysis as a "community-based and (preferably) community-driven activity involving consensus-based decisions on how to make reproducible, fair, and relevant assessments […] between new and standard catalysts". In this author's option, a complete benchmarking protocol should contain (1) a definition of all relevant test input parameters and environmental conditions; (2) a test procedure, i.e., the sequence of measurements to be performed; (3) a concise definition of the test output parameters and their evaluation criteria; (4) a well-defined and readily accessible gold standard. The current state toward a complete protocol is depicted in Figure 2 for materials-centered research and device-centered research on the OER.

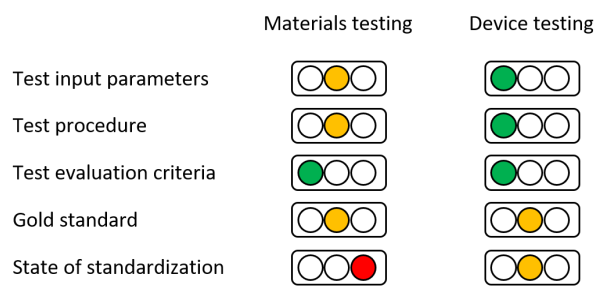

**Figure 2. State of a complete protocol and its standardization for materials and device testing.** Green (left) indicates sufficient definition, yellow (middle) indicates further work needed and red (right) indicates a lack of the criterion.

Several protocols have been published for *materials*-centered investigations of the OER. The protocol reported by McCrory et al.[4] is most widely used, yet still only by an insignificant fraction of all OER publications. Other protocols were reported by Spanos et al.,[5] Burke-Stevens et al.,[6] Peugeot,[7] and Creel et al.,[8] as well as recommendations by Wei et al.[9] (Table S1). The published protocols and also the majority of materials-centered OER studies use potential sweeps (cyclic or linear sweep voltammetry) for conditioning, surface area determination (note the pitfalls[10]) and activity determination. The protocols include either additional current and/or potential steps for activity determination or alternative current and/or potential steps for this purpose.

For *device*-centered investigations of the OER testing, Malkow et al.[11] published EU harmonized protocols for testing of low temperature water electrolyzers that employ galvanostatic sweeps *or* a list of current density setpoints (Table S1). A similar protocol was used in a round-robin study.[12] Higher current densities are included but the range of current densities overlaps with that of the materials-centered investigations. Additional definitions and experimental parameters are published in a series of reports by Tsotridis and Pilenga[13,14] that are partly based on definitions of the International Electrochemical Commission (IEC) such as standard IEC TS 60050-485:2020.

The test input parameters are defined in protocols for both materials and device testing with sufficient detail for reproduction. More environmental conditions are controlled for device testing. It will improve materials testing to mandate control of environmental conditions such as the temperature, which is readily available through jacketed electrochemical cells. In the materials-centered protocols, it is not specified how the electrochemical data is sampled, i.e., whether the current/potential reading occurs at the end of the sampling interval or by integration. This can drastically affect the contribution of capacitive currents in sweep measurement or short pulses on (desirable) high surface area materials, which would lead to an overestimation of activity metric based on electrochemical current.

The procedures vary for materials-centered testing, where potential sweeps as well as potential and current holds are performed in various combinations. The author expects that this is the main issue that reduces comparability among the protocols because different surfaces can be formed by sweep and potential/current holds[15] and because the range of potential sweeps may affect the measured currents.[16] Wei et al.[9] and Malkow et al.[11] recommend either sweeps or holds. This author recommends several current holds, i.e., a Tafel plot (note the pitfalls[17]), with increasing current density until 2 V is reached and sufficient duration to ensure a steady-state of the double layer *and* electrocatalyst microstructure (electronic structure, phase and morphology). Using current holds for both materials- and device-centered investigations to determine the activity metric(s) can be a small step towards crossing the "Valley of Death"[18] between fundamental and applied research.

Output parameters and evaluation criteria are clearly defined in the previous reports for both materials- and device-centered investigations. Common activity metrics are various (over)potentials at fixed current (density) or current (densities) at fixed (over)potential where the current is normalized by a property of the used electrocatalyst material (e.g., electrocatalyst mass or surface area) or a property of the electrode (e.g., electrode area). In addition to electrochemical data, this author urges to also report a measure of the evolved oxygen or the Faradaic efficiency.[4,5] The focus on specific activity metrics and reporting recommendations differ in details but several protocols include Tafel plots as recommended above, from which a desired metric could be calculated, most readily if the electrochemical data was published openly and FAIR[19] in a data repository.

Gold standard materials for the OER are Ni-Fe oxides in alkaline and $RuO_2$ as well as $IrO_2$ in acid. Unfortunately, the outcome of their test evaluation criteria depends strongly on details of synthesis, possible non-electrochemical post-treatment steps as well as electrochemical conditioning steps. For Ni-Fe oxide, a simple synthesis has been reported.[6] Powders of these oxides and membrane electrode assemblies (MEA) based on iridium-ruthenium oxide are also available commercially. Issues with preparation aside, there is no standard electrocatalyst or electrode consistently used in all reported protocols. Ideally, the field would need a benchmark akin to the international prototype of the kilogram (IPK) and its exact copies, which would enable to compare the reported protocols. In the field of photovoltaics, testing centers such as the European Solar Test Installation (ETSI) have been established where one sends samples for standardized tests, thus eliminating the considerable variation observed in round robin tests (on electrolyzers).[12] A clearly defined gold standard and standardized testing, especially in specialized facilities, would significantly improve reliability of reported OER activity metrics to benchmark electrocatalysts and electrodes, to identify structure-property relationships and to harness big data analysis in electrocatalysis.

In summary, the state of standardization of materials-centered investigations of the OER is less advanced as compared to device-centered investigations, yet there are no international formalized standard such as the ones that exist in corrosion science, e.g., ASTM G150-18 or DIN EN ISO 17864:2008-07, for either community. For materials testing, there are additionally no harmonized protocol or no round robin studies on gold standards. Furthermore, most reports of "highly active" materials unfortunately do not follow any of the reported protocols to obtain their activity metric(s). As pointed out by Bligaard[3] benchmarking must be a community-driven effort. Which raises the questions which are the relevant communities and should we thrive to identify a universal protocol? This author believes that the conditioning part of the protocol should be defined by sub communities, e.g., catalyst ink investigations, epitaxial thin films, alkaline electrolyzers etc. As recommended above, a current step protocol could better connect Tafel plots in materials- and device-centered OER investigations. Additional measurements could be performed after the Tafel plot or on separate samples, e.g., measurements of the Faradaic efficiency as in ref 4. These recommendations (Table S1) should be seen as a seed for the needed discussion in the community rather than competition to previous protocols. Implementing a harmonized base protocol and gold standard would be comparably little effort with large gain for the community toward truly benchmarking the OER being important in many contexts beyond water electrolysis (Figure 1).

**Acknowledgements**

This project has received funding from the European Research Council (ERC) under the European Union's Horizon 2020 research and innovation program under grant agreement No. 804092.

**Competing interests**

The author declares no competing interests

## Supplemental information

**Table 1.** Benchmarking protocols for the oxygen evolution reaction on materials and devices.

| Protocol | Step 1 | Step 2 | Step 3 | Step 4 | Step 5 | Step 6 | Step 8 | Ref |
|---|---|---|---|---|---|---|---|---|
| McCrory 2013 | CV DL, 100 mV range centered around OCP, no rotation | EIS, no rotation | CV (1.23 to 1.82 V vs RHE) at 10 mV/s, 1600 rpm, O2 | Potential steps (1.23 to 1.82 V vs RHE, 30s), 1600 rpm, O2 | Current steps (0.1 to 20 mA/cm2, 30s), 1600 rpm, O2 | 2h at 10 mA/cm2, 1600 rpm, O2 | n/a | 4 |
| Spanos 2017 A | OCP, LSV to start position of conditioning | Condition the catalyst | EIS at OCP, LSV from OCP to 1.2 V vs RHE, 5 mV/s | LSV 1.2 to 1.7 V vs RHE, 5 mV/s | Passing 1 C charge by CP or CA, LSV OCP to 1.8 V vs RHE, 5 mV/s | 2h at 10 mA/cm2 with ICP-OES analysis, flow rate of 0.86 mL/min | repeat steps 3-5 | 5 |
| Spanos 2017 B | OCP, LSV to start position of conditioning | Condition the catalyst | EIS at OCP, LSV from OCP to 1.2 V vs RHE, 5 mV/s | LSV 1.2 to 1.7 V vs RHE, 5 mV/s | Passing 1 C charge by CP or CA, LSV OCP to 1.8 V vs RHE, 5 mV/s | 2h at 1.8 V vs RHE with ICP-OES analysis, flow rate of 0.86 mL/min | repeat steps 3-5 | 5 |
| Burke 2017 | CV (0.93 to 1.68 V vs RHE), 10 mV/s | EIS at 1.53 V vs. RHE | current steps from 0.01 mA/cm2 to 10 mA/2 for 3 min | current steps from 10 mA/2 to 0.01 mA/cm2 for 3 min | repeat step 1-2 | CA at 1.53 for 1 h | n/a | 6 |
| Malkov 2018 A PEMWE | 0.1 A/cm2, 5 min | linear current sweep from 1 mA/cm2 to 2000 mA/cm2 at 0.080 A/cm2 per min, 2 V cutoff | n/a | n/a | n/a | n/a | n/a | 11 |
| Malkov 2018 A AWE + AEMWE | 0.1 A/cm2, 5 min | linear current sweep from 0.2 mA/cm2 to 400 mA/cm2 at 0.016 A/cm2 per min, 2 V cutoff | n/a | n/a | n/a | n/a | n/a | 11 |
| Malkov 2018 B PEMWE | 0.1 A/cm2, 5 min | current steps from 1 mA/cm2 to 2000 mA/cm2, with 30 s dwell and 30 s acquisition per step, 2 V cutoff | n/a | n/a | n/a | n/a | n/a | 11 |
| Malkov 2018 B AWE + AEMWE | 0.1 A/cm2, 5 min | current steps from 0.2 mA/cm2 to 400 mA/cm2, with 30 s dwell and 30 s acquisition per step, 2 V cutoff | n/a | n/a | n/a | n/a | n/a | 11 |
| Bender 2019 | 0.2 A/cm2 for 30 min, 1 A/cm2 for 30 min, 1.7 V until variation less than 1% per h | 0.0 to 0.1 A/cm2 in 0.02 A/cm2 steps and 0.2 A/cm2 steps above until 2V are reached, 5 min, optional: EIS at all steps | reverse of step 2 | optional: OCV | n/a | n/a | n/a | 12 |
| Wei 2019 A | O2 bubbling for 10-30 min | CV (1.0 to 1.7 V vs RHE) at 10 mV/s, 1600 rpm, O2 | n/a | n/a | n/a | n/a | n/a | 9 |
| Wei 2019 B | O2 bubbling for 10-30 min | CA with at least 5 potentials, for which j < 2.5 mA/cm2 | n/a | n/a | n/a | n/a | n/a | 9 |
| Peugeot 2021 | LSV at 10 mV/s until steady response | 0, 5, 10, 25, 50, 100 mA/cm2 for 5 min, extension of duration if potential was not stable | 50 mA/cm2 for 30 min | n/a | n/a | n/a | n/a | 7 |
| Creel 2022 | Condition the catalyst | LSV (1.4 V to 2.2 V vs RHE) at 10 mV/s, inert gas | EIS at OCP, OCP -50 mV, OCP +50 mV | CV DL (step 1 of McCrory 2013) | 0 mA for 3 s, 20 mA for 1 h, 0 A for 1s | n/a | n/a | 8 |
| Risch 2023 | Condition the catalyst according to (sub) community | 0 to 50 mA/cm2 in 5 mA/cm2 steps, 100 mA/cm2 and 100 mA/cm2 steps above until 2V are reached, 30 s dwell, 30 s acquisition, optional: EIS at all steps | reverse of step 2 | Optional: EIS or surface area measurement | n/a | n/a | n/a | This work |

Conditioning steps highlighted in blue, activity in green;